\documentclass[aps,prb,reprint,superscriptaddress,amsmath,amssymb]{revtex4-2}

\usepackage{graphicx}
\usepackage{bm}
\usepackage[colorlinks,allcolors=blue]{hyperref}

\begin{document}

\newcommand{\papertitle}{Accurate Microstructure-Dependent Formulas for Effective Permittivity of Two-Dimensional Plasmonic Metamaterials}
\title{\papertitle}

\author{Jaeuk Kim}
\email{pcbg304@kaist.ac.kr}
\affiliation{Department of Chemical and Biomolecular Engineering, Korea Advanced Institute of Science and Technology (KAIST), Daejeon, Republic of Korea}
\affiliation{GIST InnoCORE AI-Nano Convergence Institute for Early Detection of Neurodegenerative Diseases, Gwangju Institute of Science and Technology, Gwangju, Republic of Korea}

\author{Murray Skolnick}
\affiliation{Princeton Institute for the Science and Technology of Materials,
Princeton University, Princeton, New Jersey 08544, USA}

\author{Shin-Hyun Kim}
\email{kim.sh@kaist.ac.kr}
\affiliation{Department of Chemical and Biomolecular Engineering, Korea Advanced Institute of Science and Technology (KAIST), Daejeon, Republic of Korea}
\affiliation{GIST InnoCORE AI-Nano Convergence Institute for Early Detection of Neurodegenerative Diseases, Gwangju Institute of Science and Technology, Gwangju, Republic of Korea}

\author{Salvatore Torquato}
\email{torquato@princeton.edu}
\affiliation{Department of Chemistry, Princeton University, Princeton, New Jersey, 08544, USA}
\affiliation{Department of Physics, Princeton University, Princeton, New Jersey, 08544, USA}
\affiliation{Princeton Institute for the Science and Technology of Materials,
Princeton University, Princeton, New Jersey 08544, USA}
\affiliation{Program in Applied and Computational Mathematics,
Princeton University, Princeton, New Jersey 08544, USA}

\date{\today}

\begin{abstract}

Plasmonic metamaterials have attracted considerable attention for decades owing to their potential applications in energy harvesting, nanolasing, and biosensing.
However, designing such artificial materials with desired properties remains challenging because the primary approaches to estimating their properties, such as experiments and full-wave simulations, are costly.
An alternative is to use effective-medium approximations, but traditional formulas, such as the Maxwell-Garnett and Bruggeman approximations, are often inaccurate because they neglect geometric information.
Here, we consider Torquato's strong-contrast approximation for the effective static permittivity, which incorporates microstructural information up to four-point correlation functions.
We show that this approximation accurately predicts the complex-valued effective permittivity of two-dimensional (2D) plasmonic metamaterials over a much wider range of wavelengths, particle area fractions, and geometric models than traditional approximations.
We verify its accuracy by comparing its predictions with results from finite-difference time-domain (FDTD) simulations for five distinct models of 2D periodic plasmonic metamaterials consisting of silver particles in a dielectric matrix.
These findings establish this approximation as a reliable tool for predicting the properties of plasmonic systems in which the particles do not form large clusters.
This approximation can also be applied to disordered systems because its microstructural parameters can be computed for periodic and aperiodic media alike.
Thus, the strong-contrast approximation can facilitate the discovery of plasmonic materials with desired optical properties.

\end{abstract}

\maketitle

\section{Introduction}
\label{sec:intro}

Plasmonic metamaterials, composites of metallic nanostructures embedded in a dielectric matrix, support localized surface plasmon resonances (LSPRs) that can confine light well below the diffraction limit and strongly enhance local fields~\cite{maier_plasmonics_2007,kravets_plasmonic_2018,yang_plasmonic_2025}.
Engineering such resonances by tuning the shape, size, and arrangement of the metallic particles makes plasmonic metamaterials suitable for energy harvesting~\cite{hao_high_2010,wang_metamaterialplasmonic_2012}, nanolasing~\cite{noginov_demonstration_2009,ma_applications_2019}, and biosensing~\cite{liu_planar_2010,sreekanth_extreme_2016,wang_plasmonic_2023}.
Designing such materials therefore requires prediction of the optical properties of many candidate microstructures.
However, estimating these properties directly by experiments or full-wave simulations, such as finite-difference or finite-element methods, is costly when many candidates must be examined.
In particular, the cost of full-wave simulations escalates for plasmonic systems because each simulation must resolve the nanometer-scale features of metallic structures~\cite{boddeti_spectral_2019,zeng_inverse_2021,hemayat_integrating_2024}.
Accurate formulas that predict the effective optical properties from the microstructure are therefore of both fundamental and practical importance.

An alternative to these direct estimates is to use effective-medium approximations, which describe the composite as a homogeneous medium with an effective permittivity $\varepsilon_e(\omega)$ at frequency $\omega$~\cite{markel_introduction_2016,etrich_effective_2014}.
For simplicity, we assume (i) the \emph{quasistatic regime}, in which the radiation wavelength $\lambda$ is much longer than the characteristic length scales of the composite; and (ii) a two-phase composite in $d$ dimensions consisting of metallic particles (phase 2) with complex-valued, frequency-dependent permittivity $\varepsilon_2(\omega)$ and volume fraction $\phi_2$, embedded in a dielectric matrix (phase 1) with permittivity $\varepsilon_1$ and volume fraction $\phi_1 = 1 - \phi_2$.
Under these conditions, two of the most widely used effective-medium approximations are the Maxwell-Garnett and Bruggeman approximations.
The \emph{Maxwell-Garnett approximation}~\cite{garnett_colours_1904,sihvola_electromagnetic_1999,markel_introduction_2016} treats the metallic particles as isolated inclusions embedded in the dielectric matrix:
\begin{align} \label{eq:MG}
  \frac{\varepsilon_e (\omega) - \varepsilon_1}{\varepsilon_e (\omega) + (d-1) \varepsilon_1}
  = \phi_2 \beta_{21}(\omega),
\end{align}
where $\beta_{21}(\omega)$ is the dielectric polarizability of the metallic phase relative to the dielectric matrix, defined as
\begin{align} \label{eq:beta}
  \beta_{21}(\omega) \equiv \frac{\varepsilon_2(\omega) - \varepsilon_1}{ \varepsilon_2(\omega) + (d-1) \varepsilon_1}.
\end{align}
The \emph{Bruggeman approximation}~\cite{bruggeman_berechnung_1935,torquato_random_2002} instead treats the two phases symmetrically, with inclusions of each phase embedded in the effective medium:
\begin{align} \label{eq:BG}
  \phi_1 \frac{\varepsilon_1 - \varepsilon_e (\omega)}{\varepsilon_1 + (d-1) \varepsilon_e (\omega)}
  + \phi_2 \frac{\varepsilon_2 (\omega) - \varepsilon_e (\omega)}{\varepsilon_2 (\omega) + (d-1) \varepsilon_e (\omega)} = 0.
\end{align}
Both formulas \eqref{eq:MG} and \eqref{eq:BG} are mean-field approximations that take the reference phase to be phase 1 and the effective medium, respectively.
They depend on the microstructure only through the volume fraction $\phi_2$ and neglect strong multipolar interactions between neighboring metallic particles.
Hence, they cannot account for the shapes and arrangements of the particles.
As a result, these approximations are accurate only over a limited range of volume fractions and particle shapes, and they often fail for plasmonic composites~\cite{etrich_effective_2014,gutierrez_evaluation_2017,costa_limits_2019,maurice_maxwell_2023}.

To obtain the exact effective permittivity beyond the mean-field approximations given in Eqs.~\eqref{eq:MG} and \eqref{eq:BG}, one must account for an infinite set of $n$-point correlation functions that fully characterize the composite~\cite{milton_theory_2002,sahimi_heterogeneous_2003,torquato_random_2002}.
The \emph{strong-contrast expansion}~\cite{torquato_effective_1985,sen_effective_1989,torquato_random_2002,skolnick_effective_2025,skolnick_accurate_2025} provides a systematic way to do so.
It expresses the inverse polarizability of the effective medium relative to the dielectric matrix as a power series in the polarizability $\beta_{21}$ of the metallic phase (phase 2), in which the $n$th-order term involves correlation functions up to the $n$-point level.
Because the complete set of correlation functions is practically unobtainable, strong-contrast approximations are constructed by truncating the series after a few low-order terms.
The second-order strong-contrast approximation reproduces the Maxwell-Garnett approximation, Eq.~\eqref{eq:MG}.
The third-order approximation corrects it through the three-point parameter $\zeta_2$, and the fourth-order approximation adds a further correction through the four-point parameter $\gamma_2$.
The third- and fourth-order strong-contrast approximations accurately predict the effective permittivities of dielectric composites~\cite{torquato_effective_1985,torquato_random_2002,skolnick_effective_2025}, and their nonlocal extensions remain accurate beyond the quasistatic regime~\cite{torquato_nonlocal_2021,kim_theoretical_2024}.

For plasmonic composites, Maurice et al.~\cite{maurice_maxwell_2023} compared the Maxwell-Garnett approximation and the third-order static strong-contrast approximation, which requires only $\zeta_2$, with boundary-element simulations of three-dimensional (3D) simple cubic lattices of gold or silver nanospheres for $0.10 \leq \phi_2 \leq 0.40$.
They found that the strong-contrast approximation predicts the LSPR more accurately than the Maxwell-Garnett approximation for all values of $\phi_2$ considered.
However, this verification was restricted to one geometric model.
It thus remains unclear how the accuracy of the strong-contrast approximation depends on morphology, especially when neighboring particles have sharp corners or are separated by narrow gaps, where the local field is strongly enhanced.

\begin{figure}
  \includegraphics[width=0.45\textwidth]{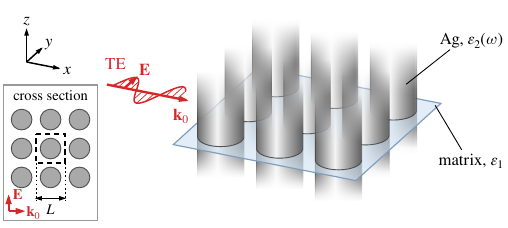}
  \caption{
  Schematic of the 2D periodic plasmonic metamaterials considered in this work.
  Each 2D model represents an array of infinitely long metallic (silver) cylinders in three dimensions.
  A plane wave with vacuum wavenumber $k_0$ is incident normal to the cylinder axes with transverse electric (TE) polarization, i.e., the electric field lies in the cross-sectional plane.
  \label{fig:schematic}
  }
\end{figure}

In this work, we scrutinize the accuracy of the fourth-order static strong-contrast approximation for two-dimensional (2D) periodic plasmonic metamaterials over a range of area fractions and particle morphologies.
These 2D models represent arrays of infinitely long metallic cylinders illuminated by transverse electric (TE) waves, whose electric field is perpendicular to the cylinder axes; see Fig.~\ref{fig:schematic}.
Specifically, we consider square-lattice arrays of silver particles with five shapes: disks, squares, diamonds, square tetramers, and diamond tetramers, as shown in Fig.~\ref{fig:geometry}.
We obtain accurate estimates of $\varepsilon_e$ from finite-difference time-domain (FDTD) simulations, using a source-driven homogenization scheme within a single unit cell~\cite{silveirinha_metamaterial_2007,costa_finitedifference_2009,fietz_currentdriven_2010}. For the strong-contrast approximation, we evaluate $\zeta_2$ by using a Monte Carlo (MC) algorithm accelerated by graphics processing units (GPUs), developed by Skolnick and Torquato~\cite{skolnick_effective_2025}.
We find that the strong-contrast approximation agrees well with the FDTD results over a much wider range of wavelengths and area fractions than the Maxwell-Garnett and Bruggeman approximations do.
It is even more accurate when neighboring particles are well separated or face one another with flat sides, whereas it becomes less accurate when sharp tips face one another across narrow gaps.

As $\zeta_2$ can be computed for any microstructure, ordered or disordered~\cite{skolnick_effective_2025}, the strong-contrast approximation also yields predictions for disordered metamaterials, which we present in Sec.~\ref{sec:comparison}.
Such disordered metamaterials are considered as attractive alternatives to periodic ones because they can be fabricated over large areas by bottom-up techniques, such as the random deposition of colloidal nanoparticles~\cite{moreau_controlledreflectance_2012,lalanne_disordered_2025}, and some of their resonances survive disorder~\cite{helgert_effective_2009}.
Engineering correlated disorder, e.g., hyperuniformity, further controls the scattering and emission of light~\cite{castro-lopez_reciprocal_2017,vynck_light_2023,yu_engineered_2021} and can make surface-enhanced Raman scattering reproducible over centimeter scales~\cite{derosa_hyperuniform_2015}.
The strong-contrast approximation therefore offers an efficient and reliable tool for discovering both ordered and disordered plasmonic metamaterials, provided that the metallic particles do not form large clusters.

The rest of this paper is organized as follows.
In Sec.~\ref{sec:methods}, we describe the strong-contrast approximation and the geometric models, and outline the MC algorithm for $\zeta_2$ and the source-driven homogenization method.
Section~\ref{sec:results} compares the predictions of the strong-contrast, Maxwell-Garnett, and Bruggeman approximations with the FDTD results.
In Sec.~\ref{sec:comparison}, we apply the strong-contrast approximation to general 2D microstructures, including the five square-lattice models alongside hexagonal arrays of disks~\cite{mcphedran_bounds_1981,torquato_random_2002} and four disordered models~\cite{skolnick_effective_2025} from earlier works.
Conclusions and an outlook are given in Sec.~\ref{sec:conclusions}.

\section{Models and Methods}
\label{sec:methods}

\subsection{Correlation Functions} \label{sec:corr-function}

A two-phase medium is a domain of $d$-dimensional Euclidean space $\mathbb{R}^d$ that is partitioned into two disjoint regions: phase 1 of volume fraction $\phi_1$ and phase 2 of volume fraction $\phi_2$~\cite{torquato_random_2002}.
For a given realization, the region $\mathcal{V}_i$ occupied by phase $i$ is specified by the phase indicator function $\mathcal{I}^{(i)}(\mathbf{x})$, which equals unity if $\mathbf{x} \in \mathcal{V}_i$ and zero otherwise.
The $n$-point correlation function of phase $i$ is defined as~\cite{torquato_random_2002}
\begin{align} \label{eq:Sn}
  S_n^{(i)}(\mathbf{x}_1, \ldots, \mathbf{x}_n) \equiv \left\langle \prod_{j=1}^{n} \mathcal{I}^{(i)}(\mathbf{x}_j) \right\rangle,
\end{align}
where angular brackets denote an ensemble average over realizations.
It gives the probability of finding $n$ points at positions $\mathbf{x}_1, \ldots, \mathbf{x}_n$ simultaneously in phase $i$.
For statistically homogeneous media, $S_n^{(i)}$ is invariant under translations, so that $S_1^{(i)}(\mathbf{x}_1) = \phi_i$ and $S_2^{(i)}(\mathbf{x}_1, \mathbf{x}_2) = S_2^{(i)}(\mathbf{r})$ with $\mathbf{r} = \mathbf{x}_2 - \mathbf{x}_1$.

\subsection{Strong-Contrast Approximation} \label{sec:SCA}

We briefly review the strong-contrast approximation considered in this work, assuming statistically isotropic two-phase materials in $d$ dimensions.
This approximation was derived by truncating the strong-contrast expansion for the effective static permittivity (or electric conductivity) at the fourth-order term~\cite{sen_effective_1989,torquato_random_2002,skolnick_effective_2025}:
\begin{align} \label{eq:SCA-4th-general}
\phi_2 \beta_{21} \frac{\varepsilon_e + (d-1) \varepsilon_1}{\varepsilon_e - \varepsilon_1}
 = 1 - (d-1)\phi_1 (\zeta_2 \beta_{21}^2 + \gamma_2  \beta_{21}^3 ),
\end{align}
where $\zeta_2$~\cite{milton_bounds_1981,torquato_effective_1985,torquato_random_2002} and $\gamma_2$~\cite{torquato_effective_1985,sen_effective_1989} are the three- and four-point parameters of the metallic phase (phase 2) defined, respectively, as
\begin{widetext}
\begin{align}
  \zeta_2 =& \frac{d^2}{(d-1)\phi_1\phi_2\,\Omega_d^2}
  \int \frac{\mathrm{d}\mathbf{x}_2\,\mathrm{d}\mathbf{x}_3}{r_{12}^d\,r_{23}^d}
  \left( d\,c_1^2 - 1 \right)\left[ S_3(1,2,3) - \frac{S_2(1,2)\,S_2(2,3)}{\phi_2} \right], \label{eq:zeta2}\\
  \gamma_2 =& \frac{d^3}{(d-1)\phi_1\phi_2\,\Omega_d^3}
  \int \frac{\mathrm{d}\mathbf{x}_2\,\mathrm{d}\mathbf{x}_3\,\mathrm{d}\mathbf{x}_4}{r_{12}^d\,r_{23}^d\,r_{34}^d} \left[ d^2 c_1 c_2 c_3 - d\,(c_1^2 + c_2^2 + c_3^2) + 2 \right] \nonumber\\
  &\times \bigg[ S_4(1,2,3,4) - \frac{S_2(1,2)\,S_3(2,3,4)}{\phi_2} - \frac{S_3(1,2,3)\,S_2(3,4)}{\phi_2} + \frac{S_2(1,2)\,S_2(2,3)\,S_2(3,4)}{\phi_2^2} \bigg], \label{eq:gamma2}
\end{align}
where $S_n(1,\ldots,n)$ is a shorthand notation for $S_n^{(2)}(\mathbf{x}_1,\ldots,\mathbf{x}_n)$; $r_{ij} = |\mathbf{x}_j - \mathbf{x}_i|$ and $\hat{\mathbf{r}}_{ij} = (\mathbf{x}_j - \mathbf{x}_i)/r_{ij}$; $c_1 = \hat{\mathbf{r}}_{12}\cdot\hat{\mathbf{r}}_{23}$, $c_2 = \hat{\mathbf{r}}_{23}\cdot\hat{\mathbf{r}}_{34}$, and $c_3 = \hat{\mathbf{r}}_{34}\cdot\hat{\mathbf{r}}_{12}$ are the cosines of the angles between pairs of separation vectors; and $\Omega_d = 2\pi^{d/2}/\Gamma(d/2)$ is the total solid angle in $d$ dimensions.
\end{widetext}

In two dimensions ($d=2$), $\gamma_2 = 0$ for any statistically isotropic medium~\cite{torquato_effective_1985,torquato_random_2002}, and thus the fourth-order approximation given in Eq.~\eqref{eq:SCA-4th-general} reduces to
\begin{align} \label{eq:SCA-4th-2D}
  \frac{\varepsilon_e - \varepsilon_1}{\varepsilon_e + \varepsilon_1}
  = \frac{\phi_2 \beta_{21}}{1 - \phi_1 \zeta_2 \beta_{21}^2}.
\end{align}
Although Eqs.~\eqref{eq:MG}, \eqref{eq:BG}, and \eqref{eq:SCA-4th-2D} are derived for static permittivities, the \emph{correspondence principle}~\cite{schulgasser_bounds_1976} allows them to be applied in the quasistatic regime with complex-valued, frequency-dependent permittivities, because their field equations take the same functional form as in the lossless static case.
For plasmonic metamaterials, we thus evaluate these approximations using the complex permittivity $\varepsilon_2(\omega)$ of the metallic phase.

\begin{figure*}
  \includegraphics[width=\textwidth]{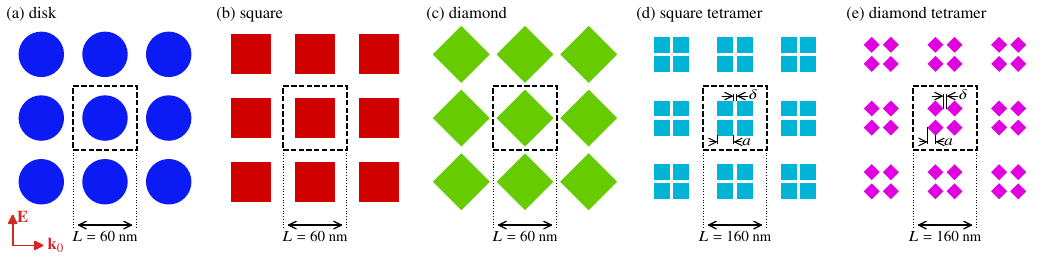}
  \caption{Representative images of the five models of 2D periodic metamaterials considered here. The models are square-lattice arrays with lattice constant $L$ and five particle shapes (colored as in Fig.~\ref{fig:zeta2}): (a) a circular disk, (b) a square, (c) a diamond, (d) a square tetramer, and (e) a diamond tetramer.
  In the tetramer models, neighboring particles are separated by a gap of $\delta = 2$~nm.
  The fundamental unit cell of each array is depicted as a dashed box, and the particles therein are chosen to have square symmetry, which guarantees that the effective permittivities are isotropic.
  We assume that electromagnetic waves are incident along one of the symmetry axes with TE polarization, as indicated by the red arrows (electric field $\mathbf{E}$ and wave vector $\mathbf{k}_0$).
  \label{fig:geometry}}
\end{figure*}

\subsection{Periodic Models}
\label{sec:models}

Figure~\ref{fig:geometry} shows the five models of 2D periodic metamaterials considered here.
Because the approximations in Eqs.~\eqref{eq:MG}, \eqref{eq:BG}, and \eqref{eq:SCA-4th-2D} assume an isotropic effective permittivity, we construct all models as square-lattice arrays whose unit cells have square symmetry such that their effective permittivities are isotropic.
The first three models [Fig.~\ref{fig:geometry}(a)--(c)] have a lattice constant $L = 60$~nm and one particle per unit cell: a circular disk, a square, and a diamond, respectively.
The last two models [Fig.~\ref{fig:geometry}(d) and (e)] have a larger lattice constant, $L = 160$~nm, and four squares or four diamonds per unit cell, separated within each unit cell by gaps of $\delta = 2$~nm.
We refer to each model by its particles---disks, squares, diamonds, square tetramers, and diamond tetramers---and to the first three and last two collectively as the \emph{single-particle} and \emph{tetramer} models.

We consider different ranges of the area fraction $\phi_2$ of the metallic phase for the five square-lattice models.
For the single-particle models, we vary $\phi_2$ from a small value up to just below the value at which neighboring particles touch.
For the tetramer models, the fixed gaps keep neighboring particles close to one another at every $\phi_2$.
We include the tetramer models because the field enhancement within narrow gaps is crucial for biosensing applications~\cite{xu_electromagnetic_2000,sundaramurthy_toward_2005}.
Because the metallic particles remain disconnected in all models, each model supports LSPRs at every $\phi_2$ considered.

\subsection{Evaluation of the Three-Point Parameter} \label{sec:zeta2}

\begin{figure}
  \includegraphics[width=0.45\textwidth]{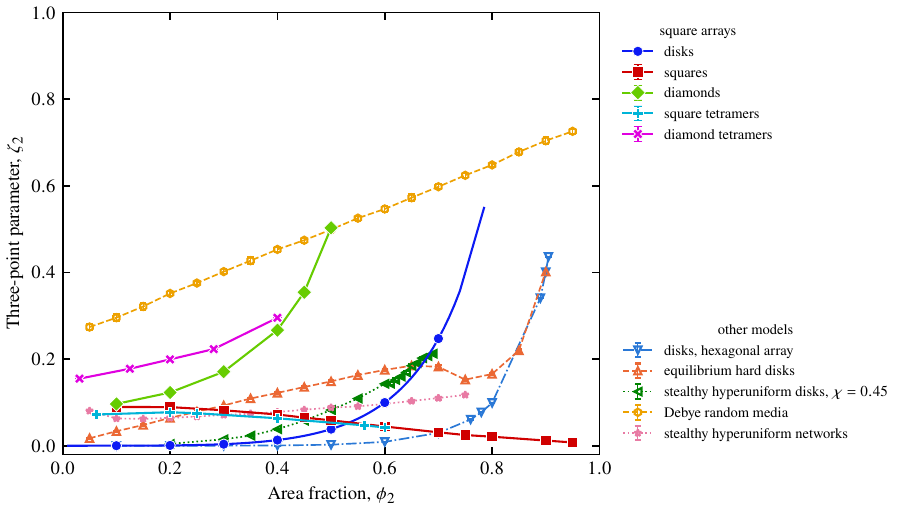}
  \caption{Three-point parameter $\zeta_2$ versus the area fraction $\phi_2$ of the metallic phase for various 2D models. Results for the five square-lattice models of Fig.~\ref{fig:geometry} (simulated with FDTD) are shown as solid lines with filled symbols; those for five other models (with $\zeta_2$ from earlier works) are shown as broken lines with open symbols: hexagonal arrays of disks~\cite{mcphedran_bounds_1981,torquato_random_2002}, equilibrium hard disks, Debye random media, stealthy hyperuniform disks with $\chi = 0.45$, and stealthy hyperuniform networks~\cite{skolnick_effective_2025}.
  For the stealthy hyperuniform networks, phases 1 and 2 of Ref.~\cite{skolnick_effective_2025} are interchanged, so that phase 2 consists of the cells enclosed by the walls and is disconnected.
  \label{fig:zeta2}}
\end{figure}

We compute $\zeta_2$ with the GPU-parallelized MC integration algorithm of Ref.~\cite{skolnick_effective_2025}.
Each MC sample is a triangle with vertices $\mathbf{x}$, $\mathbf{x}+\mathbf{r}$, and $\mathbf{x}+\mathbf{s}$, and the integral in Eq.~\eqref{eq:zeta2} is estimated as the average, over many such triangles, of the product of the phase indicator function at the three vertices, weighted by $d\,c_1^2 - 1 = \cos 2\theta$ in 2D, where $\theta$ is the angle between $\mathbf{r}$ and $\mathbf{s}$.
The term $S_2 S_2/\phi_2$ in Eq.~\eqref{eq:zeta2} does not depend on $\theta$, so its angular integral vanishes.
Three techniques reduce the statistical error~\cite{skolnick_effective_2025}: stratified sampling, which allocates the vertices $\mathbf{x}$ to square subdomains of the computational domain in proportion to the variance of the integrand there; log-uniform sampling of $r = |\mathbf{r}|$ and $s = |\mathbf{s}|$ between $r_\mathrm{min}$ and $r_\mathrm{max}$ to cancel the diverging factors $1/r$ and $1/s$; and a low-discrepancy (golden-ratio) sequence for $\theta$.
Because each subdomain is handled by a separate GPU thread, very large sample sizes are feasible.

For the four models other than the disks, we use $2^{44} \approx 1.8\times10^{13}$ samples per estimate, with $r_\mathrm{min} = 10^{-5}w$ and $r_\mathrm{max} = 600\,w$ (and the same limits for $s$), where $w$ is the unit-cell area divided by the total perimeter of the particles in the cell.
We sample the matrix phase, which gives $\zeta_1$, and use $\zeta_2 = 1 - \zeta_1$~\cite{torquato_random_2002}.
Each value of $\zeta_2$ is the mean of five runs with different random seeds, and its error bar is the standard error of the mean; the error bars are smaller than the symbols in Fig.~\ref{fig:zeta2}.
For the disks, we evaluate $\zeta_2$ from the lattice-sum series of Helsing~\cite{helsing_bounds_1994}, which reproduces the values tabulated by McPhedran and Milton~\cite{mcphedran_bounds_1981,torquato_random_2002}.

The computed values of $\zeta_2$ for the five square-lattice models of Fig.~\ref{fig:geometry} are shown in Fig.~\ref{fig:zeta2}.
At a given $\phi_2$, microstructures in which phase 2 is more connected or ramified tend to have larger $\zeta_2$~\cite{skolnick_effective_2025}.
Among the five square-lattice models, the diamonds and diamond tetramers have the largest $\zeta_2$ for $\phi_2 \le 0.40$.

In addition to the five square-lattice models, we include values of $\zeta_2$ reported in earlier works for five other models, one ordered and four disordered~\cite{mcphedran_bounds_1981,torquato_random_2002,skolnick_effective_2025}.
The periodic one is the \emph{hexagonal array of disks}~\cite{mcphedran_bounds_1981,torquato_random_2002}.
\emph{Equilibrium hard disks} are equilibrium (Gibbs) ensembles of identical nonoverlapping disks~\cite{skolnick_effective_2025}.
\emph{Debye random media} are phase-inversion-symmetric disordered media defined entirely by the autocovariance function $S_2(r) - \phi_2^2 = \phi_1 \phi_2 e^{-r/a}$, where $a$ is a characteristic length~\cite{debye_scattering_1957,torquato_random_2002}.
\emph{Stealthy hyperuniform disks} are nonoverlapping disks whose centers form disordered stealthy hyperuniform point patterns, for which the structure factor vanishes in a region around the origin of reciprocal space~\cite{uche_constraints_2004,batten_classical_2008,torquato_ensemble_2015}.
The relative size of this exclusion region is measured by the parameter $\chi$ ($\chi = 0.45$ here).
\emph{Stealthy hyperuniform networks} are obtained by decorating with walls the edges of the Delaunay-centroidal tessellations of stealthy hyperuniform point patterns with $\chi = 0.48$~\cite{skolnick_effective_2025}.
Unlike in Ref.~\cite{skolnick_effective_2025}, we take the walls to be dielectric, so that the metallic phase forms disconnected cells at every $\phi_2$; the values of $\phi_2$ and $\zeta_2$ from that reference thus become $1-\phi_2$ and $1-\zeta_2$ here.

\subsection{Full-Wave Simulations} \label{sec:simulations}

We obtain the effective permittivities of the five square-lattice models from FDTD simulations of a single unit cell, using the source-driven homogenization method~\cite{silveirinha_metamaterial_2007,costa_finitedifference_2009,fietz_currentdriven_2010}.
In this method, the unit cell $\Omega$ with Bloch-periodic boundary conditions is driven by an external current pulse with a prescribed frequency $\omega$ and Bloch wave vector $\mathbf{k}$:
\begin{align} \label{eq:source}
  \mathbf{J}_\mathrm{ext}(\mathbf{x}, t) = J_0\, \hat{\mathbf{e}}\, e^{i\mathbf{k}\cdot\mathbf{x}}\, g(t),
\end{align}
where $\hat{\mathbf{e}}$ is a unit vector in the cross-sectional plane, and $g(t)$ is a pulse.
We then evaluate the temporal Fourier components of the electric field $\mathbf{E}(\mathbf{x};\omega)$ and the displacement field $\mathbf{D}(\mathbf{x};\omega) = \varepsilon(\mathbf{x};\omega)\,\mathbf{E}(\mathbf{x};\omega)$ at each frequency.
The effective permittivity tensor $\boldsymbol{\varepsilon}_e(\omega,\mathbf{k})$ is defined by
\begin{align} \label{eq:eps-source-driven}
  \big\langle \mathbf{D}\, e^{-i\mathbf{k}\cdot\mathbf{x}} \big\rangle_\Omega
  = \boldsymbol{\varepsilon}_e(\omega,\mathbf{k}) \cdot \big\langle \mathbf{E}\, e^{-i\mathbf{k}\cdot\mathbf{x}} \big\rangle_\Omega,
\end{align}
where $\langle f \rangle_\Omega \equiv |\Omega|^{-1}\int_\Omega f(\mathbf{x})\,\mathrm{d}\mathbf{x}$ denotes the average over the unit cell.
Because both $\omega$ and $\mathbf{k}$ are imposed, the method
remains applicable when the fields decay within a few unit cells, as occurs near the LSPRs. In this work, we set $\mathbf{k} = \mathbf{0}$, and thus, the exponential factors in Eqs.~\eqref{eq:source} and \eqref{eq:eps-source-driven} reduce to unity and $\boldsymbol{\varepsilon}_e(\omega)$ relates the cell averages of $\mathbf{D}$ and $\mathbf{E}$.

We implement this scheme in one unit cell of the square lattice under periodic boundary conditions with the open-source FDTD package Meep (version 1.33)~\cite{oskooi_meep_2010}, taking the matrix to be vacuum ($\varepsilon_1 = 1$).
The silver permittivity is described by the Lorentz-Drude model of Raki\'{c} et al.~\cite{rakic_optical_1998}, as implemented in the material library of Meep (see Appendix~\ref{app:Ag}).
The same permittivity is used for $\varepsilon_2(\omega)$ in the three effective-medium approximations.
The spectrum of the pulse $g(t)$ covers frequencies $477 \le \omega/2\pi \le 1193$~THz. After a minimum duration of 50 Meep time units (approximately 166.8~fs), we monitor the discrete temporal Fourier transforms of $\mathbf{E}$ and $\mathbf{D}$ at selected frequencies, and the simulation stops when their change between three successive checks, relative to the largest such change, falls below $10^{-8}$.
We then evaluate the cell averages in Eq.~\eqref{eq:eps-source-driven} and report the scalar $\varepsilon_e = [\boldsymbol{\varepsilon}_e]_{yy} = \langle D_y^{(y)}\rangle_\Omega / \langle E_y^{(y)}\rangle_\Omega$,
where the superscript denotes the direction of the current.\footnote{For all models, the two diagonal elements of the permittivity tensor agree to within a relative difference of $10^{-11}$, and the off-diagonal elements are smaller than $10^{-12}$ relative to the diagonal elements, as required by the square symmetry of the unit cells.}
The resulting $\varepsilon_e$ is complex-valued, and its positive imaginary part measures the absorption in the metallic phase.

\begin{figure*}[h!t]
  \includegraphics[width=\textwidth]{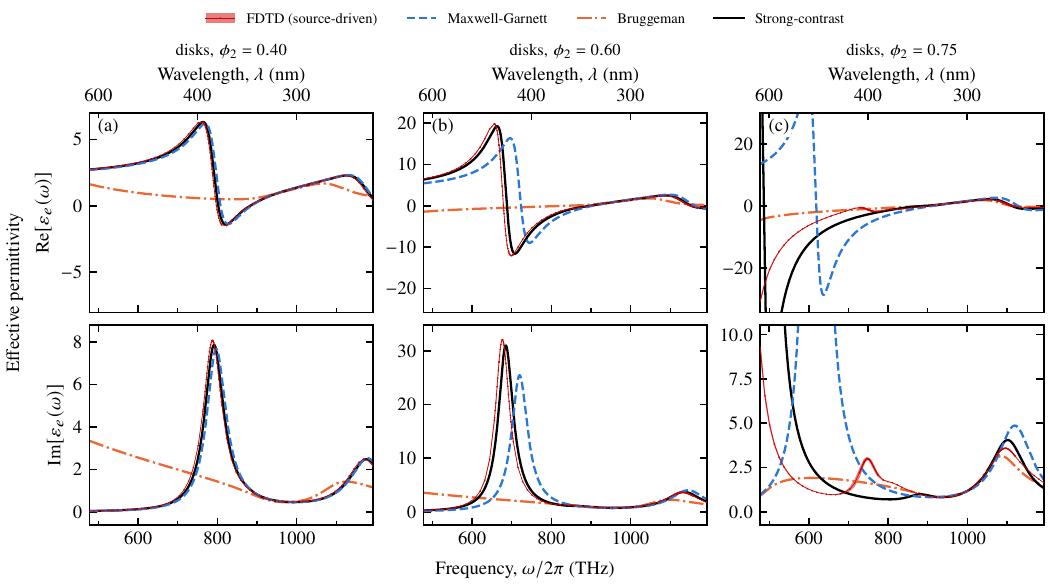}
  \caption{Comparison of the effective permittivity obtained from FDTD simulations with the predictions of three effective-medium approximations (Maxwell-Garnett, Bruggeman, and strong-contrast) for square-lattice arrays of silver disks at three area fractions: (a) $\phi_2=0.40$, (b) $\phi_2=0.60$, and (c) $\phi_2=0.75$.
  The strong-contrast approximation is in excellent agreement with the FDTD results over the entire spectral range for $\phi_2 \le 0.60$.
  \label{fig:comp-circle}}
\end{figure*}

We estimate the numerical error arising from the spatial discretization of the silver interfaces by repeating each simulation at several resolutions.
We use 4000, 5000, 6000, and 7000 pixels per micrometer for the single-particle models, and 4000, 5000, and 6000 pixels per micrometer for the tetramer models, meaning that the 2-nm gaps span 8--12 pixels.
The uncertainties $\Delta\,\mathrm{Re}\,\varepsilon_e$ and $\Delta\,\mathrm{Im}\,\varepsilon_e$ are half the spread of the respective values over these resolutions and are shown as shaded bands in Figs.~\ref{fig:comp-circle} and \ref{fig:comp-4dia} and in Figs.~\ref{fig:comp-square}--\ref{fig:comp-4squares} of the Supplemental Material.
Inside the spectral range considered, the median relative uncertainty $|\Delta\varepsilon_e|/|\varepsilon_e|$, with $|\Delta\varepsilon_e|^2 = (\Delta\,\mathrm{Re}\,\varepsilon_e)^2 + (\Delta\,\mathrm{Im}\,\varepsilon_e)^2$, is 0.7\%.
It is largest near the resonance peaks, reaching 18\% for diamonds at $\phi_2 = 0.45$, where the corners of neighboring particles are separated by gaps of 3.1~nm; see Fig.~\ref{fig:comp-diamonds}(c).

\section{Numerical Verifications}
\label{sec:results}

\begin{figure*}
  \includegraphics[width=\textwidth]{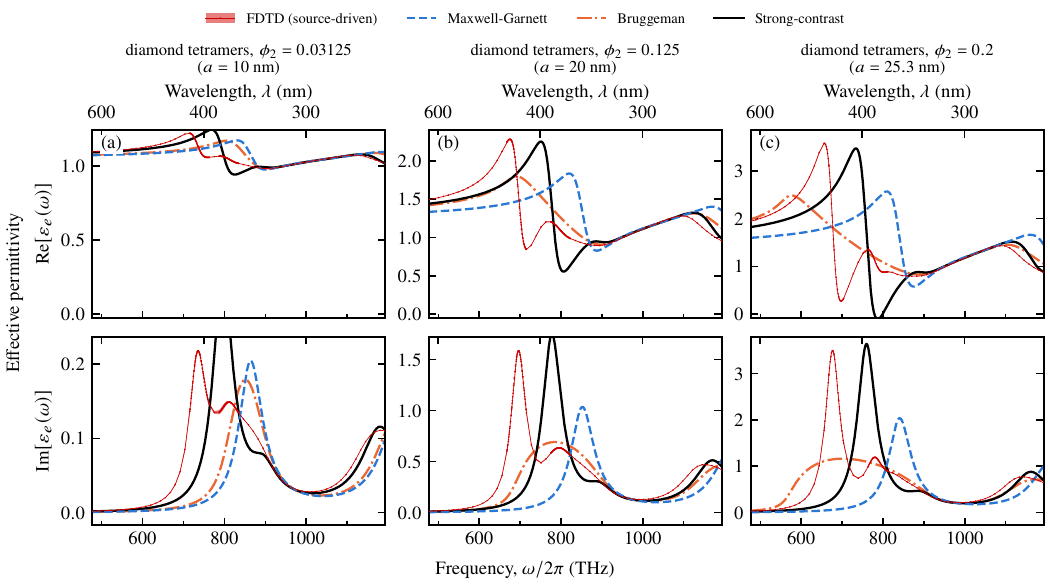}
  \caption{Comparison of the effective permittivity obtained from FDTD simulations with the predictions of three effective-medium approximations (Maxwell-Garnett, Bruggeman, and strong-contrast) for square-lattice arrays of silver diamond tetramers at three area fractions: (a) $\phi_2=0.03125$, (b) $\phi_2=0.125$, and (c) $\phi_2=0.20$.  \label{fig:comp-4dia}}
\end{figure*}

\begin{figure*}
  \includegraphics[width=\textwidth]{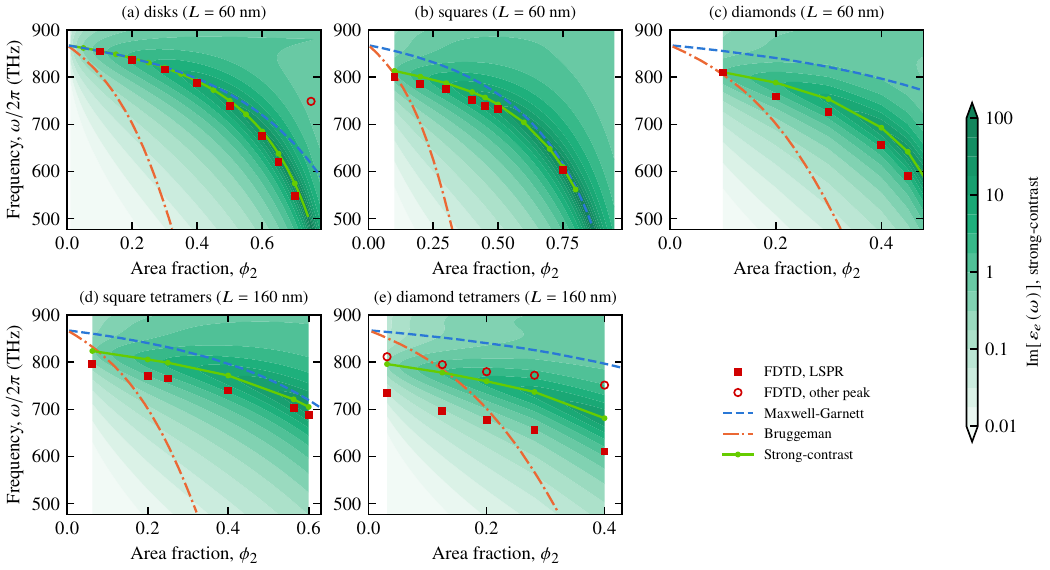}
  \caption{Contour maps of the imaginary part of the effective permittivity, $\mathrm{Im}\,\varepsilon_e$, predicted by the strong-contrast approximation as a function of the frequency $\omega/2\pi$ and area fraction $\phi_2$ for the five 2D square-lattice plasmonic metamaterials composed of silver particles: (a) disks, (b) squares, (c) diamonds, (d) square tetramers, and (e) diamond tetramers.
  The contours are drawn only over the range of $\phi_2$ for which $\zeta_2$ was computed (Fig.~\ref{fig:zeta2}), with $\zeta_2$ interpolated linearly between the computed values.
  On the contour maps, we overlay the LSPR frequencies from the FDTD results and the Maxwell-Garnett, Bruggeman, and strong-contrast approximations.
  The LSPR frequency is taken as the frequency at which $\mathrm{Im}\,\varepsilon_e$ is maximal for $500 \le \omega/2\pi \le 900$~THz, a range that contains the LSPR and excludes the interband peak; for the disks at $\phi_2 = 0.75$, the FDTD LSPR lies below this range and is not shown.
  In panels (a) and (e), the additional weak peaks (other than the LSPR and interband peaks) observed in the FDTD results are denoted by red circles.
  \label{fig:comp-all-pred}}
\end{figure*}

We investigate the accuracy of the strong-contrast approximation \eqref{eq:SCA-4th-2D} and the two other effective-medium approximations---the Maxwell-Garnett approximation \eqref{eq:MG} and the Bruggeman approximation \eqref{eq:BG}---by comparing their predictions with the FDTD results for the five square-lattice models.
We first describe the FDTD results and the predictions of the three approximations for the single-particle models and then for the tetramer models.
For brevity, we show the detailed comparisons only for the disks and the diamond tetramers (Figs.~\ref{fig:comp-circle} and \ref{fig:comp-4dia}), for which the strong-contrast approximation is the most and the least accurate, respectively; for the other three models, we outline only the differences here and present the detailed comparisons in Figs.~\ref{fig:comp-square}, \ref{fig:comp-diamonds}, and \ref{fig:comp-4squares} of the Supplemental Material.
Finally, we use the contour maps of Fig.~\ref{fig:comp-all-pred} to display the LSPR frequencies of all five models across the ranges of $\phi_2$ considered and to examine their shape-dependent shift at low $\phi_2$.

For the single-particle models, the FDTD results for $\mathrm{Im}\,\varepsilon_e$ exhibit two peaks at most values of $\phi_2$ shown in Figs.~\ref{fig:comp-circle}, \ref{fig:comp-square}, and \ref{fig:comp-diamonds}.
The strong peak at lower frequency comes from the LSPR, and the secondary peak at higher frequency (hereafter the interband peak) originates from the interband term of the silver permittivity (Appendix~\ref{app:Ag}).
Both peaks redshift as $\phi_2$ increases and, consequently, the multipolar coupling between neighboring particles strengthens.

For these single-particle models, both the Maxwell-Garnett and strong-contrast approximations also yield the LSPR and interband peaks and their redshift with increasing $\phi_2$, but the latter approximation predicts the peak positions more accurately.
While the Maxwell-Garnett approximation predicts the redshift of the LSPR peak due to dipolar coupling~\footnote{In the lossless limit, the 2D Maxwell-Garnett approximation, Eq.~\eqref{eq:MG}, resonates when $\phi_2 \beta_{21} = 1$, i.e., at $\varepsilon_2 = -\varepsilon_1 (1+\phi_2)/(1-\phi_2)$; this value becomes more negative as $\phi_2$ increases, so the resonance moves to a lower frequency. This condition arises from the mean dipolar field of the neighboring particles. For a square array of cylinders, this mean field is exact at dipole order, and the corrections from higher multipoles start at order $\phi_2^4$~\cite{rayleigh_influence_1892,perrins_transport_1979}.}, the strong-contrast approximation also partly accounts for the higher-order multipolar coupling through $\zeta_2$.
Specifically, for the disks, the Maxwell-Garnett and strong-contrast approximations agree well with the FDTD results up to $\phi_2 = 0.40$ and $0.60$, respectively, above which each increasingly overestimates the LSPR frequency [Figs.~\ref{fig:comp-circle} and \ref{fig:comp-all-pred}(a)].
At $\phi_2 = 0.75$, where the LSPR peak lies below the plotted spectral range, the FDTD results show an additional peak at $\omega/2\pi \approx 750$~THz, which none of the three approximations predicts [Fig.~\ref{fig:comp-circle}(c)].
Among the single-particle models, both the Maxwell-Garnett and strong-contrast approximations become less accurate as $\phi_2$ increases, except for the squares (Figs.~\ref{fig:comp-square} and \ref{fig:comp-diamonds}).
For the squares, both approximations instead become more accurate and converge to the permittivity of silver as $\phi_2$ approaches unity [Fig.~\ref{fig:comp-square}(c)]. Although the Bruggeman approximation reproduces the interband peak for the squares and diamonds, it is the least accurate of the three for these single-particle models: it deviates strongly from the FDTD results at all three area fractions in Fig.~\ref{fig:comp-circle}, and even at $\phi_2 = 0.20$ (not shown), its LSPR peak for the disks lies at a 16\% lower frequency than the peak in the simulations.

For the tetramer models, the FDTD results show one LSPR peak and one interband peak, as in the single-particle models.
The diamond tetramers, however, exhibit an additional weak peak near the LSPR peak (Fig.~\ref{fig:comp-4dia}).
For both tetramer models, the Maxwell-Garnett and strong-contrast approximations predict the LSPR and interband peaks, and the strong-contrast approximation is more accurate at every $\phi_2$.
Among the five models, the error of the strong-contrast approximation in the LSPR frequency at a given $\phi_2$ is largest for the diamond tetramers.
For the square tetramers, the Bruggeman approximation predicts both peaks at the lowest area fraction, $\phi_2 = 0.0625$, but its accuracy degrades immediately as $\phi_2$ increases, and it is already unreliable at $\phi_2 = 0.25$ (Fig.~\ref{fig:comp-4squares}).

The contour maps of $\mathrm{Im}\,\varepsilon_e$ from the strong-contrast approximation as a function of frequency and $\phi_2$ illustrate its accuracy and limitations in predicting the LSPR frequency; see Fig.~\ref{fig:comp-all-pred}.
For each model, because $\zeta_2$ varies continuously with $\phi_2$, the LSPR peaks form a continuous curve (shown in green).
On these contour maps, we overlay the predictions of the Maxwell-Garnett and Bruggeman approximations together with the FDTD results for comparison.
For the Maxwell-Garnett and Bruggeman approximations, the LSPR frequency is independent of particle shape and converges in the dilute limit ($\phi_2 \to 0$) to the isolated-cylinder value, $\omega/2\pi \approx 866$~THz, where $|\varepsilon_2(\omega) + \varepsilon_1|$ is minimized.
Consequently, they cannot capture the effects of particle shape on the shift in the LSPR frequency.
At the lowest $\phi_2$ considered for each model, however, the FDTD LSPR frequency lies 5--8\% below the Maxwell-Garnett prediction for the squares, diamonds, and square tetramers, and 15\% below it for the diamond tetramers.
The strong-contrast approximation closes 76\%, 98\%, 57\%, and 53\% of this difference for the squares, diamonds, square tetramers, and diamond tetramers, respectively.
It captures this shift because $\zeta_2$ for these four models remains nonzero (0.07--0.15) at their lowest $\phi_2$, whereas $\zeta_2 < 10^{-4}$ for the disks at $\phi_2 \le 0.10$ (Fig.~\ref{fig:zeta2}).

The contour maps in Fig.~\ref{fig:comp-all-pred} also show how the accuracy of the strong-contrast approximation depends on the particle morphology.
At the same area fraction, it becomes less accurate when particles face one another across narrow gaps.
Specifically, at $\phi_2 = 0.20$, its LSPR frequency deviates from the FDTD results by 1.9\% and 3.9\% for the squares and diamonds but by 4.5\% and 12.1\% for the corresponding square and diamond tetramers.
For the same gap width, the approximation is also more accurate when flat sides, rather than sharp tips, face one another across the gaps, as shown by the comparison between the square tetramers [Fig.~\ref{fig:comp-all-pred}(d)] and the diamond tetramers [Fig.~\ref{fig:comp-all-pred}(e)].
These two observations are consistent with the picture that the strong-contrast approximation accounts for multipolar coupling only partially and cannot fully capture the contribution of the extremely high field enhancement in such gaps.
Moreover, because Eq.~\eqref{eq:SCA-4th-2D} takes the dielectric matrix as the reference phase, it can be unreliable when the metallic phase forms large clusters or percolates~\cite{torquato_random_2002,skolnick_accurate_2025}.

\section{Predictions Across Different Models}
\label{sec:comparison}

Having established the accuracy and limitations of the strong-contrast approximation for the five square-lattice models, we now apply it to more general 2D models.
Specifically, in addition to the five square-lattice models, we consider the five other models described in Sec.~\ref{sec:zeta2}: the hexagonal arrays of disks and four disordered models.
Figure~\ref{fig:comp-all-fixed-phi2} compares the effective permittivities predicted by the strong-contrast approximation for the ten models at $\phi_2 = 0.20$, $0.40$, and $0.60$.
This comparison directly confirms that both the LSPR and interband peak frequencies predicted by the strong-contrast approximation redshift further as $\zeta_2$ increases at fixed $\phi_2$.
In addition, the value of $\mathrm{Im}\,\varepsilon_e$ at the LSPR peak, i.e., the absorption strength, grows with $\zeta_2$, reflecting the strong dependence of the LSPR on the microstructure.
At $\phi_2 = 0.40$, for instance, the LSPR shifts from $\omega/2\pi = 797$~THz ($\lambda = 376$~nm) for the hexagonal array of disks ($\zeta_2 \approx 0$) to 609~THz ($\lambda = 492$~nm) for the Debye random media ($\zeta_2 = 0.45$), while the peak value of $\mathrm{Im}\,\varepsilon_e$ increases from 7.7 to 26; see Fig.~\ref{fig:comp-all-fixed-phi2}(b).
Among these models, the Debye random media at $\phi_2 = 0.60$ [Fig.~\ref{fig:comp-all-fixed-phi2}(c)] lie above their percolation threshold, $\phi_2 = 1/2$~\cite{skolnick_accurate_2025}, so their metallic phase is connected and their predicted spectrum should be interpreted with care: for any $\zeta_2 < 1$, Eq.~\eqref{eq:SCA-4th-2D} gives a finite $\varepsilon_e$ as $|\varepsilon_2| \to \infty$ and thus cannot reproduce the low-frequency Drude-like response of a connected metal.

In this approximation, the LSPR at fixed $\phi_2$ is set by $\zeta_2$, not by disorder alone.
At $\phi_2 = 0.20$, the stealthy hyperuniform disks ($\zeta_2 = 0.004$) have nearly the same spectrum as the square and hexagonal arrays of disks ($\zeta_2 < 0.001$), with an LSPR frequency within 0.3\% of theirs.
The LSPR of the equilibrium hard disks ($\zeta_2 = 0.064$) lies between those of the ordered disk arrays and the squares ($\zeta_2 = 0.089$).
Particle shape and narrow gaps shift the LSPR to still lower frequencies: the predicted LSPRs of the diamonds ($\zeta_2 = 0.12$) and diamond tetramers ($\zeta_2 = 0.20$) occur at 789 and 760~THz, compared with 811~THz for the equilibrium hard disks.

\begin{figure*}
  \includegraphics[width=\textwidth]{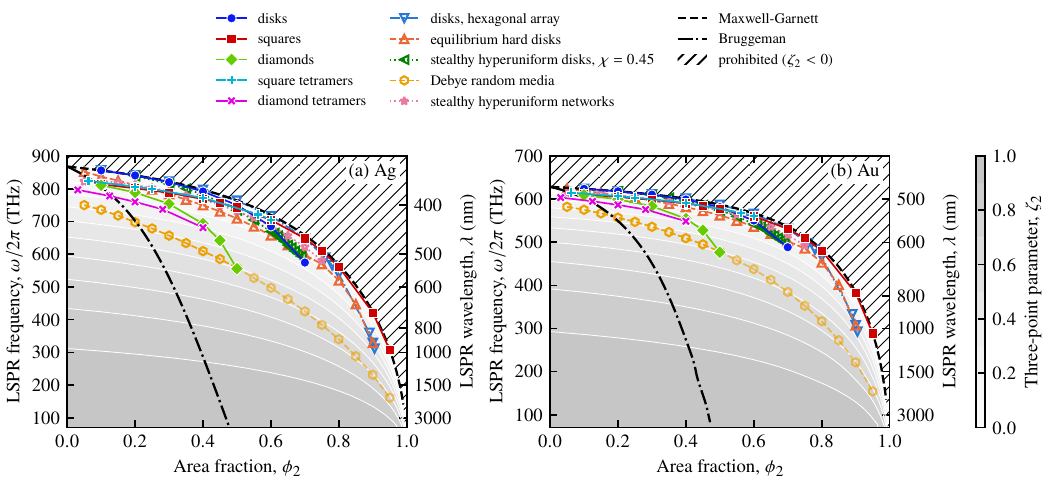}
  \caption{LSPR frequency $\omega/2\pi$ predicted by the strong-contrast approximation as a function of the area fraction $\phi_2$ for (a) silver and (b) gold particles.
  The gray shade represents the three-point parameter $\zeta_2$: each boundary between two shades traces the LSPR at a fixed value of $\zeta_2$, in steps of 0.1.
  The dashed and dash-dotted black curves show the LSPR frequencies predicted by the Maxwell-Garnett and Bruggeman approximations, respectively.
  The Maxwell-Garnett approximation coincides with the strong-contrast one at $\zeta_2 = 0$, and the hatched region above this curve corresponds to $\zeta_2 < 0$, which is prohibited in the fourth-order strong-contrast approximation because $0 \le \zeta_2 \le 1$.
  The Bruggeman curve ends just below $\phi_2 = 1/2$, above which this approximation describes a connected metallic phase without an LSPR.
  Symbols denote the LSPR frequencies of the ten models from Fig.~\ref{fig:zeta2} at the values of $\phi_2$ and $\zeta_2$ shown there, using the same marker shapes, colors, and line styles.
  Faded symbols of the Debye random media indicate the cases with $\phi_2 > 1/2$, in which the metallic phase is percolated and the predictions should be interpreted with care.
  The LSPR is taken as the highest maximum of $\mathrm{Im}\,\varepsilon_e$ for $\omega/2\pi \le 1050$~THz ($k_0 \le 22~\mu\mathrm{m}^{-1}$), and the right axes give the corresponding vacuum wavelength.
  The permittivities of silver and gold are described in Appendix~\ref{app:Ag}.
  \label{fig:lspr-zeta2}}
\end{figure*}

Because the LSPR frequency in the strong-contrast approximation depends only on $\phi_2$ and $\zeta_2$ for given constituents, we can generate unified maps of the LSPR frequency.
Figure~\ref{fig:lspr-zeta2}(a) shows such a map for silver, in which the gray shades indicate the values of $\zeta_2$ and each contour curve between two shades traces the LSPR frequency as a function of $\phi_2$ at fixed $\zeta_2$.
The hatched region in the upper right side of the panel corresponds to $\zeta_2<0$, which is prohibited in the fourth-order strong-contrast approximation.
Because $\zeta_2$ can be estimated for any microstructure, a given 2D plasmonic microstructure, ordered or disordered, is represented as a single point on this map, as illustrated by the five square-lattice models and the five other models described in Sec.~\ref{sec:zeta2}.
The loci of the LSPR predicted by the Maxwell-Garnett and Bruggeman approximations are depicted as the dashed and dash-dotted curves, respectively.
The analogous map for gold, shown in Fig.~\ref{fig:lspr-zeta2}(b), is obtained by replacing $\varepsilon_2(\omega)$ with the model permittivity of gold described in Appendix~\ref{app:Ag}.

The maps for both metals show three important points.
First, the LSPR frequency redshifts as $\zeta_2$ increases at fixed $\phi_2$, implying that the Maxwell-Garnett curve ($\zeta_2 = 0$) is an upper bound on the LSPR frequency of any microstructure at a given $\phi_2$ within this approximation.
Second, the vertical span of the shaded region at a given $\phi_2$ indicates the range over which the LSPR can be tuned by engineering the microstructure; this range widens slightly as $\phi_2$ decreases.
Third, to design a microstructure with a target LSPR frequency, one can read off the corresponding ranges of $\phi_2$ and $\zeta_2$ from the map and thereby narrow the search to microstructures with these values, provided that their morphology satisfies the conditions under which the strong-contrast approximation is accurate.

\section{Conclusions and Discussion}
\label{sec:conclusions}

We have studied the accuracy of the fourth-order static strong-contrast approximation, given in Eq.~\eqref{eq:SCA-4th-2D}, in its application to 2D plasmonic metamaterials under TE polarization.
By comparison with the FDTD results for the five square-lattice models, we have demonstrated that the strong-contrast approximation outperforms the traditional mean-field approximations---the Maxwell-Garnett and Bruggeman approximations---in predicting the frequency-dependent $\varepsilon_e(\omega)$ and the LSPR frequency over a wide range of microstructures and $\phi_2$; see Sec.~\ref{sec:results}.
The accuracy of the strong-contrast approximation depends on the morphology.
It agrees closely with the FDTD results when neighboring particles are well separated or face one another with flat sides, as for the disks at $\phi_2 \le 0.60$ and the squares at $\phi_2 = 0.75$ and $0.95$.
By contrast, it becomes less accurate when sharp tips of metallic particles face one another across narrow gaps, as for the diamonds near $\phi_2 = 0.5$ and the diamond tetramers.
Its predictions should also be used with caution in two regimes:
(i) beyond the quasistatic regime, where the absolute length scale of the microstructure affects the plasmonic response~\cite{maier_plasmonics_2007}, whereas the approximation, derived from the static problem, is independent of this length scale; and
(ii) for a percolated metallic phase, as in the Debye random media for $\phi_2 > 1/2$, whose Drude-like response the approximation cannot describe.

Within these limits, the approximation is practical: a single value of $\zeta_2$ per microstructure, evaluated with the GPU-accelerated MC algorithm (Sec.~\ref{sec:zeta2}) or taken from earlier works~\cite{skolnick_effective_2025}, gives the full spectrum $\varepsilon_e(\omega)$ for any metal and dielectric matrix through the closed-form expression in Eq.~\eqref{eq:SCA-4th-2D} (Fig.~\ref{fig:lspr-zeta2}).
For a target LSPR frequency, precalculated maps such as those in Fig.~\ref{fig:lspr-zeta2} narrow the design space of microstructures to a range of $\phi_2$ and $\zeta_2$.
This inverse-design approach can be enhanced by a machine-learning algorithm trained on microstructures that are labeled with $\zeta_2$ and $\phi_2$ and for which the strong-contrast approximation is well suited.

Several directions remain for future work: verifying the fourth-order approximation in three dimensions, where $\gamma_2$ does not vanish; extending the approximation to anisotropic media; and testing approximations built for clustered media, such as that of Ref.~\cite{skolnick_accurate_2025}, for microstructures whose metallic phase percolates.

\begin{acknowledgments}
J.K. and S.K. were supported by the InnoCORE program of the Ministry of Science and ICT (GIST InnoCORE KH0860).
M.S. and S.T. were supported by the Army Research Office under Cooperative Agreement No. W911NF-22-2-0103.
Simulations were performed on computational resources managed and supported by the Korea Institute of Science and Technology Information (KISTI).
\end{acknowledgments}

\appendix

\section{Permittivities of Silver and Gold} \label{app:Ag}

We describe the permittivity of silver by the Lorentz-Drude model of Raki\'{c} et al.~\cite{rakic_optical_1998}, as implemented in the material library of Meep~\cite{oskooi_meep_2010}.
With the time dependence $e^{-i\omega t}$, it reads
\begin{align} \label{eq:eps-Ag}
  \varepsilon_2(\omega) = 1 - \frac{f_0\,\omega_p^2}{\omega(\omega + i\Gamma_0)} + \sum_{j=1}^{5} \frac{f_j\,\omega_p^2}{\omega_j^2 - \omega^2 - i\omega\Gamma_j},
\end{align}
where the second term describes the free electrons (Drude term) and the sum describes interband transitions by five Lorentz oscillators.
The plasma frequency is $\hbar\omega_p = 9.01$~eV, and the oscillator strengths $f_j$, resonance energies $\hbar\omega_j$, and damping energies $\hbar\Gamma_j$ are listed in Table~\ref{tab:Ag}.
Meep specifies the range of validity of this model as vacuum wavelengths from 248~nm to 12.4~$\mu$m.
The spectral range considered in this work, $477 \le \omega/2\pi \le 1193$~THz, or vacuum wavelengths of 251--628~nm, lies within this range.

For reference, the same material library describes gold by Eq.~\eqref{eq:eps-Ag} with $\hbar\omega_p = 9.03$~eV and the parameters listed in Table~\ref{tab:Ag}~\cite{rakic_optical_1998,oskooi_meep_2010}.
Meep specifies the range of validity of this model as vacuum wavelengths from 248~nm to 6.2~$\mu$m.

\begin{table}
  \caption{Parameters of the Lorentz-Drude model, Eq.~\eqref{eq:eps-Ag}, for silver (considered in this work, $\hbar\omega_p = 9.01$~eV) and gold ($\hbar\omega_p = 9.03$~eV), as implemented in the material library of Meep~\cite{oskooi_meep_2010}. The values are taken from Ref.~\cite{rakic_optical_1998}. The $j = 0$ row is the Drude term.
  \label{tab:Ag}}
  \begin{ruledtabular}
  \begin{tabular}{ccccccc}
    & \multicolumn{3}{c}{Silver (Ag)} & \multicolumn{3}{c}{Gold (Au)} \\
    \cline{2-4} \cline{5-7}
    $j$ & $f_j$ & $\hbar\omega_j$ (eV) & $\hbar\Gamma_j$ (eV) & $f_j$ & $\hbar\omega_j$ (eV) & $\hbar\Gamma_j$ (eV) \\
    \hline
    0 & 0.845 & 0 & 0.048 & 0.760 & 0 & 0.053 \\
    1 & 0.065 & 0.816 & 3.886 & 0.024 & 0.415 & 0.241 \\
    2 & 0.124 & 4.481 & 0.452 & 0.010 & 0.830 & 0.345 \\
    3 & 0.011 & 8.185 & 0.065 & 0.071 & 2.969 & 0.870 \\
    4 & 0.840 & 9.083 & 0.916 & 0.601 & 4.304 & 2.494 \\
    5 & 5.646 & 20.29 & 2.419 & 4.384 & 13.32 & 2.214 \\
  \end{tabular}
  \end{ruledtabular}
\end{table}

\clearpage
\onecolumngrid
\setcounter{page}{1}
\renewcommand{\thepage}{S\arabic{page}}
\setcounter{section}{0}
\setcounter{figure}{0}
\setcounter{table}{0}
\setcounter{equation}{0}
\renewcommand{\thesection}{S\arabic{section}}
\renewcommand{\thefigure}{S\arabic{figure}}
\renewcommand{\thetable}{S\arabic{table}}
\renewcommand{\theequation}{S\arabic{equation}}
\renewcommand{\theHsection}{S\arabic{section}}
\renewcommand{\theHfigure}{S\arabic{figure}}
\renewcommand{\theHtable}{S\arabic{table}}
\renewcommand{\theHequation}{S\arabic{equation}}

\begin{center}
  {\large\bfseries Supplemental Material for\\[2pt]
  ``\papertitle''}\\[10pt]
  Jaeuk Kim, Murray Skolnick, Shin-Hyun Kim, and Salvatore Torquato
\end{center}
\suppressfloats[t] \vspace{1em}

\section{Numerical Verifications for the Other Three Models} \label{apx:additional}

\begin{figure*}
  \includegraphics[width=\textwidth]{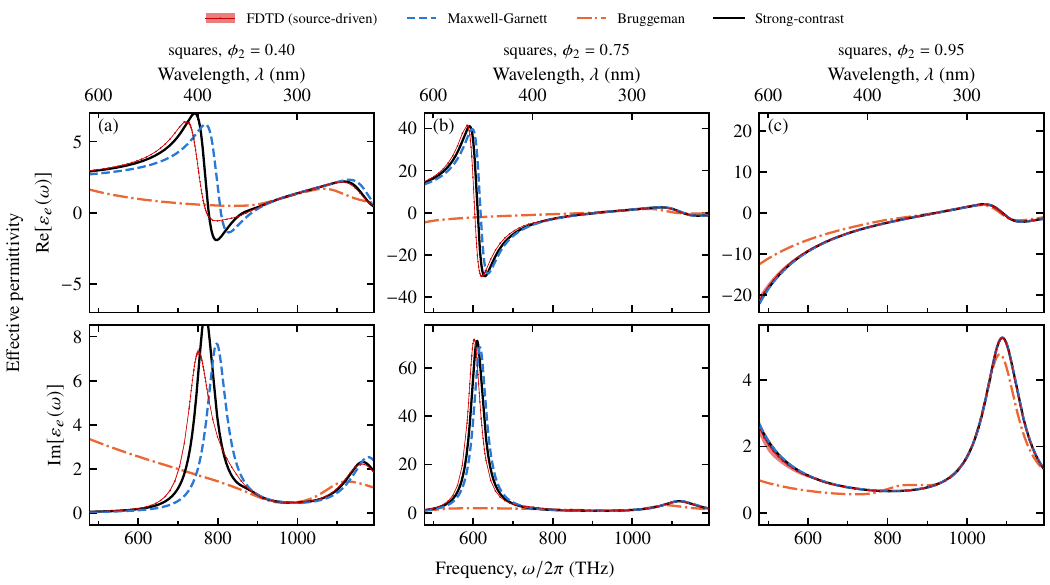}
  \caption{Comparison of the effective permittivity obtained from FDTD simulations with the predictions of three effective-medium approximations (Maxwell-Garnett, Bruggeman, and strong-contrast) for square-lattice arrays of silver squares at three area fractions: (a) $\phi_2=0.40$, (b) $\phi_2=0.75$, and (c) $\phi_2=0.95$.  \label{fig:comp-square}}
\end{figure*}

Here, we elaborate on the comparisons, outlined in the main text for brevity, of three effective-medium approximations---Maxwell-Garnett (MGA) \eqref{eq:MG}, Bruggeman (BGA) \eqref{eq:BG}, and strong-contrast (SCA) \eqref{eq:SCA-4th-2D}---with the FDTD results for the other three models (squares, diamonds, and square tetramers).

\begin{figure*}
  \includegraphics[width=\textwidth]{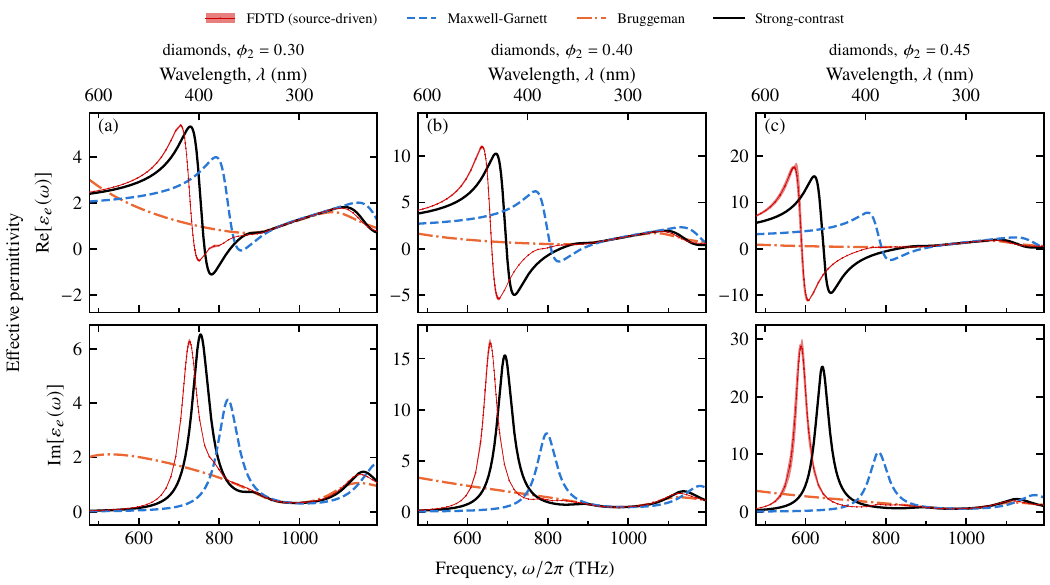}
  \caption{Comparison of the effective permittivity obtained from FDTD simulations with the predictions of three effective-medium approximations (Maxwell-Garnett, Bruggeman, and strong-contrast) for square-lattice arrays of silver diamonds at three area fractions: (a) $\phi_2=0.30$, (b) $\phi_2=0.40$, and (c) $\phi_2=0.45$. \label{fig:comp-diamonds}}
\end{figure*}

Figure~\ref{fig:comp-square} shows the estimated effective permittivities for the squares at three values of the area fraction $\phi_2$.
At $\phi_2=0.40$ [Fig.~\ref{fig:comp-square}(a)], the FDTD results show two peaks: the localized surface plasmon resonance (LSPR) peak at $\omega/2\pi \approx 750$~THz and a weaker peak at $\omega/2\pi \approx 1160$~THz, which originates from the interband term of the permittivity model of silver (Appendix~\ref{app:Ag}) and is henceforth called the interband peak.
Both MGA and SCA capture these two peaks, but MGA overestimates the peak frequencies more than SCA does.
BGA captures the interband peak within the spectral range of interest, but its predictions are otherwise very inaccurate.
The same relative ranking holds at $\phi_2=0.75$ [Fig.~\ref{fig:comp-square}(b)]: both MGA and SCA accurately predict the permittivity response, and MGA overestimates the LSPR frequency slightly more than SCA does.
At $\phi_2=0.95$ [Fig.~\ref{fig:comp-square}(c)], the two approximations are indistinguishable and agree with the FDTD results.
Thus, SCA predicts the resonance frequencies more accurately than MGA across $\phi_2$, but both converge to the permittivity of silver, $\varepsilon_2(\omega)$, as $\phi_2$ tends to unity.
Counterintuitively, the two approximations become more accurate as $\phi_2$ increases for the squares, whereas for the other single-particle models, namely, the disks in Fig.~\ref{fig:comp-circle} and the diamonds in Fig.~\ref{fig:comp-diamonds}, the deviations from the FDTD results generally increase with $\phi_2$ because of the stronger interactions between neighboring particles.
This behavior is consistent with the fact that $\zeta_2$ of the squares decreases toward zero as $\phi_2$ approaches unity; see Fig.~\ref{fig:zeta2}.

We make the analogous comparisons for the diamonds at $\phi_2=0.30$, $0.40$, and $0.45$; see Fig.~\ref{fig:comp-diamonds}.
The case of $\phi_2=0.50$---also known as the 2D periodic checkerboard, in which neighboring diamonds touch at their corners---is excluded because the FDTD results do not converge as the resolution increases.
The FDTD results at each value of $\phi_2$ show an LSPR peak and an interband peak.
Both MGA and SCA predict these two peaks at higher frequencies; the peak positions predicted by SCA are closer to the FDTD results than those predicted by MGA.
In contrast to the squares in Fig.~\ref{fig:comp-square}, the discrepancies between these two approximations and the FDTD results increase as $\phi_2$ increases.
This trend is expected: as $\phi_2$ increases, the gaps between the facing corners of neighboring diamonds narrow (from 13.5~nm at $\phi_2=0.30$ to 3.1~nm at $\phi_2=0.45$), which strengthens the interaction between neighboring particles and thus shifts the actual resonance further toward lower frequencies.
Again, BGA is the least accurate of the three approximations, although it does predict the interband peak within the spectral range of interest.

\begin{figure*}
  \includegraphics[width=\textwidth]{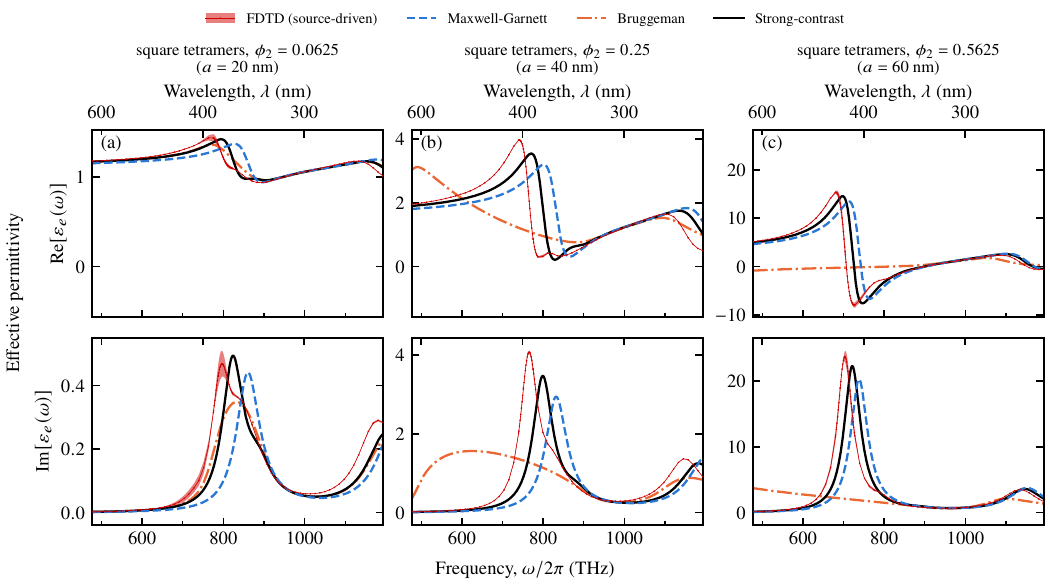}
  \caption{Comparison of the effective permittivity obtained from FDTD simulations with the predictions of three effective-medium approximations (Maxwell-Garnett, Bruggeman, and strong-contrast) for square-lattice arrays of silver square tetramers at three area fractions: (a) $\phi_2=0.0625$, (b) $\phi_2=0.25$, and (c) $\phi_2=0.5625$.  \label{fig:comp-4squares}}
\end{figure*}

We also compare the three approximations with the FDTD results for the square tetramers at three area fractions, $\phi_2=0.0625$, $0.25$, and $0.5625$, which correspond to particle side lengths $a=20$, $40$, and $60$~nm, respectively; see Fig.~\ref{fig:comp-4squares}.
The FDTD results show an LSPR peak and an interband peak, at lower and higher frequencies, respectively.
Both MGA and SCA capture these two peaks, and the peak positions predicted by SCA are closer to the FDTD results than those predicted by MGA.
These two approximations become slightly less accurate as $\phi_2$ increases up to around $0.25$ and then become monotonically more accurate, as confirmed by additional FDTD simulations at $\phi_2=0.20$, $0.40$, and $0.60$ (not shown).
At the lowest area fraction, $\phi_2=0.0625$, BGA predicts both the LSPR and interband peaks within the spectral range of interest.
However, its accuracy degrades notably as $\phi_2$ increases, and it is already unreliable at $\phi_2=0.25$.

\pagebreak

\section{Effective Permittivities Predicted by the Strong-Contrast Approximation}
\label{sec:ten-models}

\begin{figure*}
  \includegraphics[width=\textwidth]{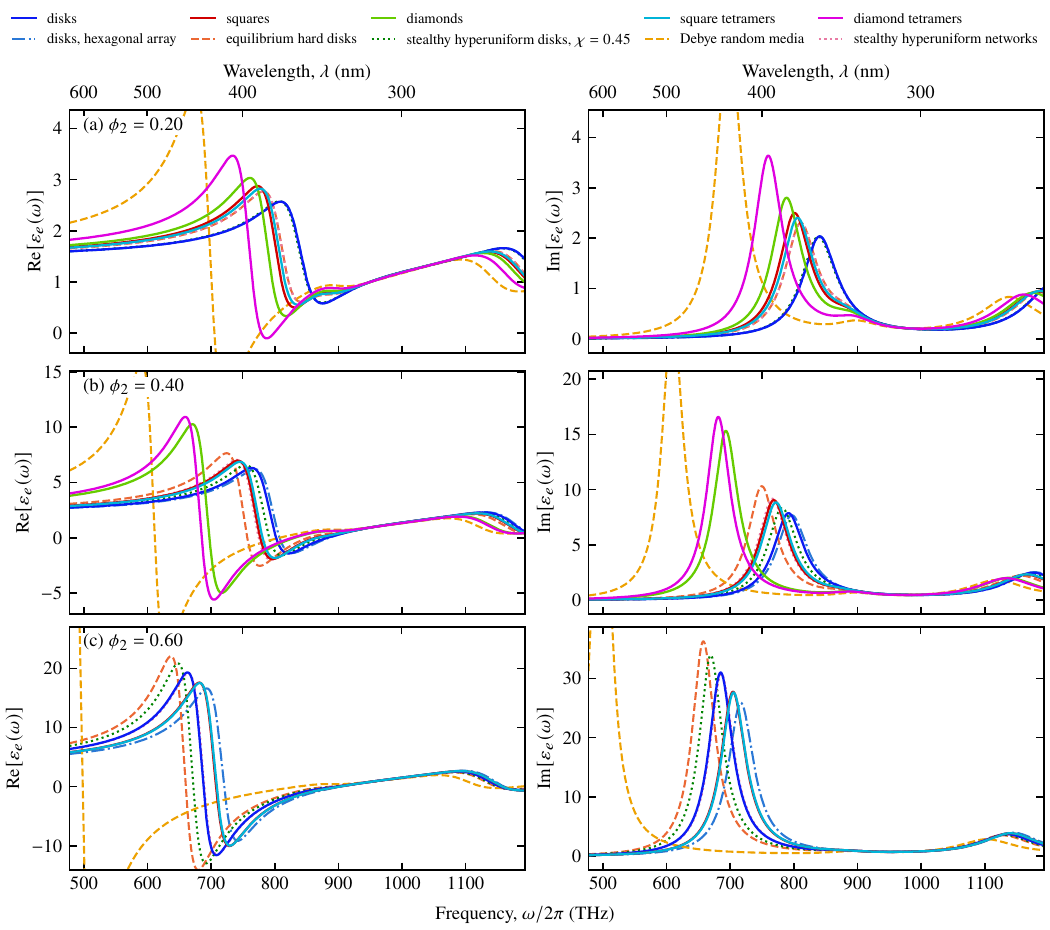}
  \caption{Comparison of the effective permittivities predicted by the strong-contrast approximation for the 2D models at three area fractions: (a) $\phi_2=0.20$, (b) $\phi_2=0.40$, and (c) $\phi_2=0.60$.
  Each panel shows all models in Fig.~\ref{fig:zeta2} for which $\zeta_2$ is available at that $\phi_2$.\label{fig:comp-all-fixed-phi2}}
\end{figure*}

Figure~\ref{fig:comp-all-fixed-phi2} shows the effective permittivities predicted by SCA, Eq.~\eqref{eq:SCA-4th-2D}, at $\phi_2 = 0.20$, $0.40$, and $0.60$ for the 2D models and $\zeta_2$ values in Fig.~\ref{fig:zeta2}.
At each $\phi_2$, the LSPR shifts to lower frequency and its peak value of $\mathrm{Im}\,\varepsilon_e$ grows as $\zeta_2$ increases.
These spectra are discussed in Sec.~\ref{sec:comparison} of the main text.

\end{document}